\documentclass[onecolumn]{rsauthor_astroph} 
\usepackage{graphicx}

\jname{Phil. Trans. R. Soc. A}

\artnum{XXXXXXXX}

\begin{document} 
\title{Supernovae and Cosmology with Future European Facilities} 
\author{I. M. Hook}
\address{University of Oxford Astrophysics, Denys Wilkinson
  Building, Keble Road, Oxford OX1 3RH, UK  and \\ INAF Osservatorio
  Astronomica di Roma, via di Frascati 33, 00040 Monte Porzio Catone,
  Italy}

\keywords{supernovae, future facilities}
\date{}

\maketitle

\begin{abstract}

 Prospects for future supernova surveys are discussed, focusing on the
 ESA Euclid mission and the European Extremely Large Telescope
 (E-ELT), both expected to be in operation around the turn of the
 decade. Euclid is a 1.2m space survey telescope that will operate at
 visible and near-infrared wavelengths, and has the potential to find
 and obtain multi-band lightcurves for thousands of distant
 supernovae. The E-ELT is a planned general-purpose ground-based
 40m-class optical-IR telescope with adaptive optics built in, which
 will be capable of obtaining spectra of Type Ia supernovae to
 redshifts of at least four. The contribution to supernova cosmology
 with these facilities will be discussed in the context of other
 future supernova programs such as those proposed for DES, JWST, LSST
 and WFIRST.
\end{abstract}

\section{Introduction} 

The discovery of the accelerating expansion of the Universe
\cite{riess98, perlmutter99} was one of the biggest breakthroughs of
the last twenty years and was awarded the 2011 Nobel Prize for
Physics. However the fundamental question remains: what drives this
acceleration? One possibility is that it is driven by a mysterious
component of the Universe, termed ``Dark Energy'' that exerts negative
pressure and that constitutes about $70\%$ of the energy density of
the universe. Focus has now turned to measuring the equation-of-state
parameter, $w (= \rm pressure/density)$ of this Dark Energy. Current
measurements of $w$ are consistent with $-1$, the value expected if
the behaviour of Dark Energy can be described by the cosmological
constant ($\Lambda$) in Einstein's field equations. Although this
would perhaps be the simplest explanation, large problems remain, for
example there is no natural explanation for such a small and yet
non-zero value for the vacuum energy density. Other explanations such
as ``quintessence'' and modified gravity have been proposed, and
distinguishing between these models is one of the major challenges of
modern physics. A measurement of $w$ not equal to $-1$ (at any
redshift) would would rule out the cosmological constant explanation
and would have profound consequences for physics. Therefore there is
considerable effort being directed towards improving measurements of
$w$ using several techniques including weak lensing, Baryon Acoustic
Oscillations (BAO), and Type Ia Supernovae (SNe~Ia).

\begin{figure}
\centering{ \includegraphics[width=2.35in]{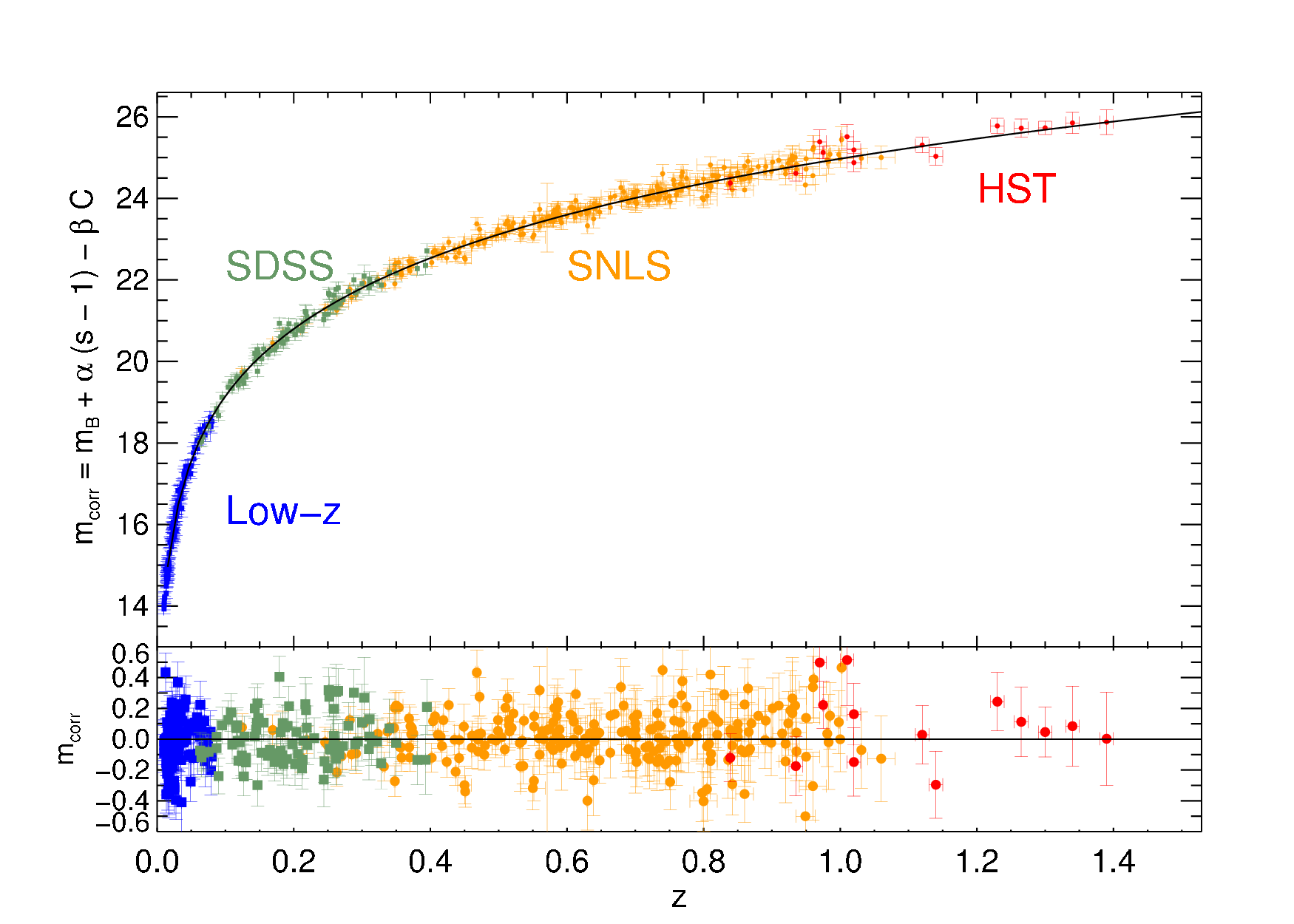}
 \includegraphics[width=2.35in]{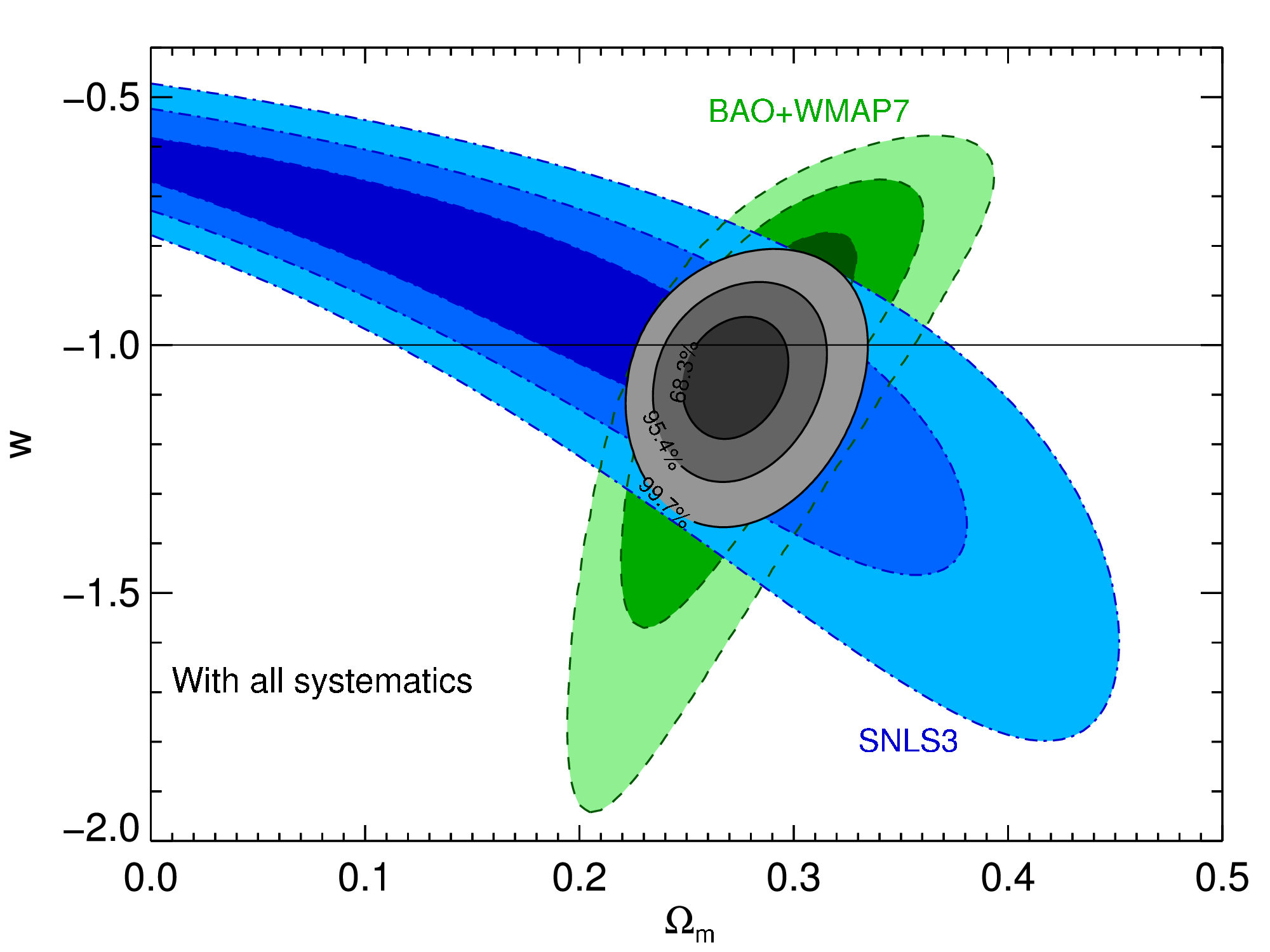}}
 \caption{SNLS 3-year results \cite{conley11, sullivan11, guy10} Left:
   (reproduced from \cite{conley11}) Hubble diagram for 472 SNe~Ia
   (123 low-z, 93 SDSS, 242 SNLS, 14 HST). Right: (reproduced from
   \cite{sullivan11}) cosmological constraints from SNe~Ia (blue),
   including systematics and assuming a flat universe. The grey
   contours show combined constraints with WMAP7 and SDSS LRG power
   spectra constraints (green) and a prior on the Hubble constant
   $H_0$ from SHOES. (Online version in colour)}
\label{snls3}
\end{figure}

The best constraints on Dark Energy to date measure $w$ consistent
with $-1$ at the $\sim 7\%$ level (including statistical and
systematic uncertainties), assuming a flat universe and a constant
equation of state \cite{sullivan11, suzuki12}, see
Fig~\ref{snls3}. Presently systematic errors are estimated to be
comparable to the statistical uncertainties in SN cosmology. The major
sources of systematic error are related to photomeric calibration,
particularly when comparing distant SNe to nearby SN samples that have
been compiled in a different way and with different rest-frame
wavelength coverage. New nearby searches are addressing this problem
(for examples, see papers by B. Schmidt and J. Tonry at this meeting).

A second source of systematic error arises from our lack of
understanding of the colours of SNe~Ia. SNe~Ia show a range of colours,
and the colour correlates inversely with brightness. Such a relation
could be caused by dust extinction in the SN host galaxies or could be
an intrinsic property of SNe~Ia themselves, or a combination of the
two. To make progress, any future SN survey should measure SNe~Ia in
multiple bands, particularly towards redder wavelengths where the
effects of dust are reduced and there is evidence for reduced
intrinsic dispersion \cite{freedman09}.

Finally, another potential systematic effect is evolution of the SN~Ia
population. We know that SN~Ia spectra are similar at low and high
redshift (up to $z\sim 0.9$) \cite{bronder08, walker11} although
differences in the in the UV have recently been seen
\cite{maguire12}. We also know that SN~Ia rates and broadband SNe~Ia
lightcurve properties are dependent on the host galaxy type. The
luminosity of SNe~Ia depends on host galaxy type even after correction
using the usual stretch and colour correction method
\cite{sullivan10}. Extending the redshift range over which SNe~Ia are
measured will allow better control of these effects. To date only a
handful of SNe~Ia have been observed at $z >1$ at any wavelength. Such
observations are extremely difficult from the ground because (a) the
SNe are faint, and (b) the peak of the SNe~Ia spectral energy
distribution moves into the NIR where the sky background is
higher. HST has therefore been the primary route for finding such SNe
\cite{riess07,amanullah10,suzuki12}.

\section{Future facilities} 

In the following sections, several upcoming projects are decribed that
will make dramatic advances in SN~Ia cosmology. These approach the
problem in two main ways. The first is improving statistics (which can
also lead to improved control of systematics through construction of
sub-samples). The second is extending the wavelength coverage of
high-redshift SN samples into the near-infrared (near-IR).

\subsection{DES and VISTA} 

The Dark Energy Survey (DES)\footnote{http://www.darkenergysurvey.org}
is an international private-public partnership involving institutions
from USA, Spain, UK, Brazil and Germany with the goal of mounting and
operating a 3 sq deg CCD camera on the CTIO Blanco 4m telescope. DES
will carry out a 5000 sq deg survey over 5 years starting in late
2012. It will survey the sky in g,r,i,z,Y bands.

The SN component of DES has been described in detail by
\cite{bernstein12}. 30 square degress will be imaged repeatedly in the
g,r,i,z bands and is expected to result in 4000 well measured SNe~Ia
in the redshift range $0.05 < z< 1.2$.  An external spectroscopy
program is being planned that will provide robust redshifts of the
host galaxies (which aid SN photometric classification and are
necessary for accurate fitting of cosmological parameters). In
addition, spectra will be obtained for a subset of the SNe themselves
($< 20\%$), some of which will have detailed multi-epoch spectroscopy.

The 4m VISTA telescope, equipped with a 67 Mpix NIR camera and
operated by ESO, has been carrying out public NIR surveys in
Z,Y,J,H,Ks bands since 2009. The SN component within the VIDEO survey
\cite{jarvis12} is expected to produce about 100 SNe~Ia with $z< 0.5$
observed in Y and J bands. These SNe will overlap with the DES sample
and will therefore have both optical and near-IR lightcurves.

\subsection{LSST}
\label{sec:lsst}
The Large Synoptic Survey Telescope
(LSST)\footnote{http://www.lsst.org} is a U.S.-based public-private
partnership that aims to construct and operate a 8.4m survey telescope
(6.7m effective diameter) equipped with a very wide field camera (9.6
sq deg) with u,g,r,i,z,y filters. The main survey will cover $\rm
\sim20000 deg^2$ every 3-4 days, and by the end of the survey each
field will have been imaged over 1000 times. The ``Deep Drilling
Fields'' will cover a smaller area but reaching a deeper limiting
magnitude per visit and with faster cadence (to be determined),
thereby producing SN lightcurves of higher quality and reaching higher
redshift than the main survey.  20-40 such fields are envisioned, of
which the locations of 4 have already been selected. LSST is due to
start operation at the end of the decade and will have a $\sim 10$
year survey lifetime.

The SNe~Ia case has been described in detail in the LSST Science book
(Chapter 11, Wood-Vasey et al). For SN Ia cosmology the key gain is
massive statistics: 50000 SNe~Ia per year are expected, with redshift
up to $\sim 0.8$ (main survey) and up to $\sim 1$ (Deep Drilling
Fields).  Such statistics also improve systematics by allowing the
sample to be split into subsets, for example depending on host galaxy
type. However because of the very large sample size, only a small
fraction will have spectroscopic redshifts and classifications. The
LSST SN survey will also allow tests on isotropy and homogeneity, and
tests of SN Ia evolution. LSST will also find core-collapse SNe and
measure SN rates (for all Types).

\subsection{JWST} 
\label{sec:jwst}

The James Webb Space Telescope
(JWST)\footnote{http://www.jwst.nasa.gov/} is a joint project between
NASA, the European Space Agency and the Canadian Space Agency. It
consists of a 6.5m space observatory with four instruments optimised
to the infrared wavelength range. Launch is currently expected in
2018. Being an observatory as opposed to a survey instrument, JWST's
science use will be determined mainly from PI programs. JWST's key
advantage for SN cosmology is sensitivity at infrared wavelengths,
allowing it to reach well above $z=1$. Several possible ways that JWST
could advance SN cosmology have been suggested.

$z > 1$ : In the white paper ``James Webb Space Telescope Studies of
Dark Energy'' \cite{gardner10} it is demonstrated that in 1 year using
1080 hours, JWST could find and follow 60 $z>1$ SNe, including
obtaining spectra with JWST itself.

$z > 2$: Riess \& Livio (2006) make the case in that observing at
$z>2$, where the effects of Dark Energy are expected to be very
modest, will provide a key test for evolution in the SN~Ia
population. They estimate that $\sim 1.5$ SNe~Ia could be found within
each NIRCAM field reaching 10nJy in K. They suggest monitoring a few
NIRCAM fields with cadence $\sim 100$ days.

$z \sim 4$: The JWST exposure time calculator predicts that JWST can
reach 2 magnitudes below peak brightness of a typical SN~Ia lightcurve
at $z=4$ in 10 000s. By monitoring 10 NIRCAM fields for 5 years, a
sample of $\sim 50$ objects could be found.  Spectroscopic
confirmation could be obtained using future ground-based extremely
large telescopes (see next section).

\subsection{The European ELT} 
\label{sec:eelt}

Three extremely large telscope projects are currently underway: the
Giant Magellan Telescope (GMT)\footnote{http://www.gmto.org/}, the
Thirty Meter Telescope (TMT)\footnote{http://tmt.org/} and the
European Extremely Large Telescope
(E-ELT)\footnote{http://www.eso.org/sci/facilities/eelt/}. These
projects are at a comparable stage of development and all aim for
first light around the turn of the decade. In this paper I focus on
the European ELT, but the broad conclusions are similar, or scalable,
to all three projects.

\begin{figure}
\centering{ \includegraphics[width=4.7in]{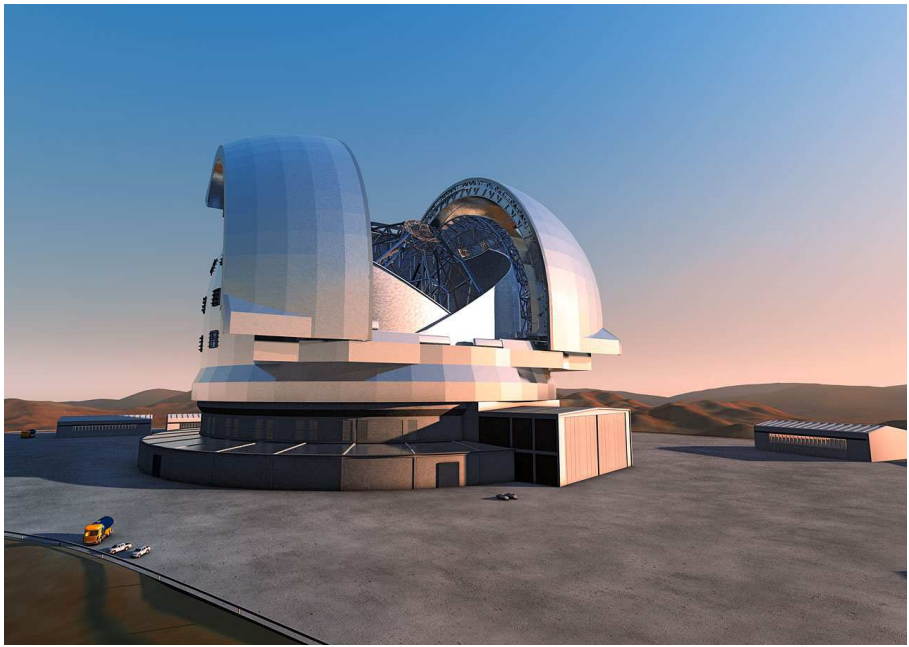}}
\caption{Artist's impression of the European Extremely Large Telescope
  (E-ELT). Image credit: ESO (Online version in colour)}
\end{figure}

The E-ELT project aims to design and construct a 39m diameter
optical-IR telescope, which will be the largest optical/IR telescope
in the World. The project is run by ESO on behalf of its member
states. The design is a novel 5 mirror concept with adaptive optics
built in, and the telescope will be situated on Cerro Armazones in
Chile. The first three instruments have been selected: a
diffraction-limited camera and an integral-field spectrograph for
first light, and a mid-IR imager-spectrograph to arrive shortly
afterwards. The full instrumentation suite will be built up over first
decade of operation.  Since the time of this Royal Society Discussion
meeting, the ESO Council has approved the E-ELT programme subject to
confirmation of funding from some member states.

The E-ELT is a general-purpose observatory and the science case is
very general \cite{elt_scicase}. In additon to its clear power for
identification of varying sources from other facilities (e.g. GRBs),
examples of time domain science with E-ELT include Solar System
observations (including study of weather and volcanic activity),
Exo-planets (incuding radial velocity, direct detection and transit
measurements) and study of the motions of stars and stellar flares in
the vicinity of the Galactic centre. Furthermore, with specialised
high-time resolution detectors, E-ELT could study extreme physics
(pulsars, neutron stars, black holes) stellar phenomena, transits and
occultations \cite{htra}.

The E-ELT will be particularly powerful for spectroscopic observations
of SNe~Ia at high redshift. Because they are point sources, SN
observations benefit from the telescope's adaptive optics
capability. Such spectroscopic observations are needed in order to
unambiguously classify the SNe and to determine their
redshifts. Simulations show that with the HARMONI integral-field
spectrograph \cite{thatte10} and using the telescope's adaptive optics
system, the SiII feature near a rest-frame wavelength of 4000\AA\ is
clearly detected even at $z=4$ (at which redshift the feature is
observed in the K-band; Hook, 2010). Since the presence of SiII and
absence of hydrogen is the defining signature of SNe~Ia, this means
that it will be possible to confirm SN Types even out to $z=4$. To
obtain E-ELT spectroscopy of a sample of 50 objects $1< z < 4$ (which
could be reasonably discovered by JWST, see section~\ref{sec:jwst})
would require of order 400 hours spread over the over 5 years of the
survey.

\subsection{Euclid} 
\label{sec:euclid}

Euclid is a 1.2m optical-IR space telescope within ESA's Cosmic Vision
2015-2025 programme \cite{laureijs11}. Its primary science goal is
precison measurement of cosmological parameters via weak lensing and
galaxy clustering techniques.  ESA's downselection process in October
2011 resulted in selection of Euclid for the second M-class launch
slot. In June 2012 the mission passed the important milestone of
adoption of the mission by ESA. Launch is currently scheduled to take
place in Q4 2019.

The instrumentation consists of an optical imager and a near infrared
imager and spectrograph, both with field of view of 0.5 sq deg. The
satellite will be launched by a Soyuz rocket and will operate at the
L2 Lagrange point for a 6 year mission duration.

Euclid's Wide survey will cover $\ge 15000$ sq deg, with imaging in a
single broad optical band (R+I+Z) to a depth of AB=24.5 (10$\sigma$
for a point source), imaging in three near infrared bands (Y, J, H) to
a depth of AB=24 (5$\sigma$, extended source) and NIR slitless
spectroscopy to a depth of $\rm 3\times 10^{-16} cm ^{-2}s^{-1}$
(3.5$\sigma$ unresolved line flux).

Additional Deep fields will cover $\ge 40$ sq deg, reaching 2
magnitudes deeper than the Wide survey in both optical and NIR imaging
and NIR Spectroscopy. The Deep fields are primarily for calibration
purposes but the $\sim 40$ repeat visits will enable a vast range of
additional science incuding detection and study of variable, moving
and transient objects.

In addition, ideas for specialised dedicated surveys for SNe and
microlensing of exoplanets are being considered, although these are
not currently in the baseline. These will be further explored as the
survey design is optimised.

For SNe~Ia cosmology, one example strategy has been developed that
would allow exploration of a new redshift range, to $z\sim 1.5$.
Simulations show that 6 months of Euclid survey time could be used to
carry out a survey of 20 sq deg with 4 day cadence. When combined with
simiultaeous ground-based observations in I and z bands, this results
in 1700 well measured SNe~Ia in the redshift range $0.75 < z < 1.5$
with measurements covering a consistent rest-frame wavelength
range. Such a survey would make a significant improvement in the
measurement of cosmological parameters from existing surveys at the
time, and would add an independent method to Euclid's primary
cosmological probes, thereby enhancing the mission's overall impact.

\subsection{WFIRST} 

The proposed NASA mission WFIRST (Wide-Field Infra-Red Survey
Telecope) emerged as the top priority large space mission in the
U.S. ``New Worlds New Horizons'' 2010 Decadal Survey. Its science
drivers are measurement of the expansion history of the Universe and
galaxy clustering (exploring Dark Energy and modified gravity theories
via the weak lensing, supernova and BAO techniques); Exoplanets (via
microlensing); Deep NIR surveys; a Galactic plane survey; High-z QSOs
and a guest observer program.

Definition of the hardware is in progress. The interim report of the
Science Definition Team study \cite{green11} described a 1.3m off-axis
telescope operating from $\rm 0.6-2.0\mu m$, and a $\sim 0.3$ sq deg
FOV covered by imaging in 5 filters and slitless spectroscopy with $R
\sim 200$ from $\rm 1.1-2.0\mu m$.  An additional $R\sim 75$ prism
would be available for SN spectroscopy. Launch is anticipated for
$\sim 2020$, assuming phase A starts in 2013.

The WFIRST SN~Ia program goals are stated as $> 100$ SNe per $\Delta z
=0.1$ redshift bin in the range $0.4<z<1.2$ per dedicated 6 months
(spread out during the mission lifetime). The goal is to achieve an
error on distance modulus of $< 0.02$ mag per $\Delta z=0.1$ redshift
bin. They also consider an `optimistic' case where the redshift range
is extended to 1.5 and systematic effects are reduced.

\section{Conclusions}

\begin{figure}
 \includegraphics[width=4.7in]{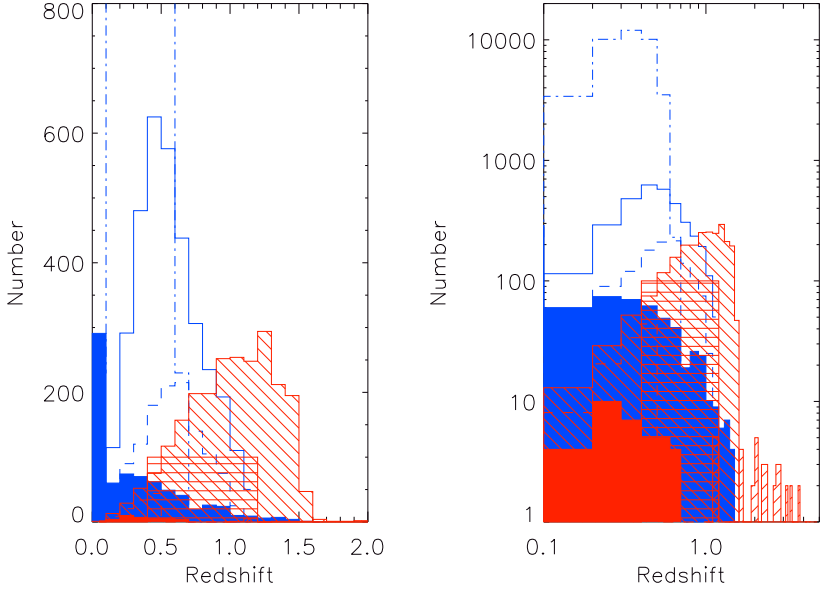}
  \caption{(Online version in colour) Redshift distributions from
    existing and planned SN surveys in (left panel) linear and (right
    panel) log space. Optical surveys are shown in blue: the existing
    Union 2.1 sample \cite{suzuki12} (filled histogram); DES (solid
    line); LSST wide survey and Deep Drilling fields (dot-dashed and
    dashed lines respectively). Note that the LSST values are for only
    one year of operation and may be a factor $\sim 10$ higher for the
    final sample. Near-IR samples are shown in red: the existing CSP
    sample \cite{freedman09} (filled histogram); VISTA sample (assumed
    to be 100 objects sampled from the DES dristribution up to z=0.5,
    dashed line); a possible Euclid survey if scheduling allows
    (hatched histogram); the WFIRST SN goals (horizontal hatched
    histogram); an example very high-redshift sample that could be
    compiled by JWST and ELTs (diagonal hatched histogram at $z>1$).}
\label{nz}
\end{figure}

There is a spectacular suite of new facilities on the horizon that
will make dramatic advances in the quantity and quality of SN samples
for cosmology. Several of these facilities are already planning SN
programs and have made detailed predictions of the achievable results
in terms of numbers of objects and in some cases cosmological
constraints (e.g. in terms of the Dark Energy Task Force Figure of
Merit). Such figure of merit calculations depend critically on the
assumptions made for systematic effects and priors, so comparison is
not straightforward and is not attempted here. However by comparing
the N(z) distributions of existing and future samples we can obtain a
simple visual impression of the gains we can expect (Figure
~\ref{nz}).

In summary, the gains likely to emerge are:
\begin{itemize}
\item an enormous gain in statistics (at least 2 orders of magnitude)
  of optical SN~Ia measurements, from a combination of current nearby
  searches, DES and then LSST
\item an enormous gain (2 orders of magnitude) in numbers of
  observed-frame NIR observations of distant SNe, from VISTA and then
  space-based surveys with Euclid (survey strategy permitting) and
  WFIRST, potentially allowing extension of the SN~Ia Hubble diagram
  up to $z\sim 1.5$.
\item better control of systematic effects, arising from (a) the ability to create
  sub-samples from large parent samples, and (b) from the improved
  wavelength coverage resulting from combined optical-NIR observations
  of the same SNe 
\item extension of the SN~Ia Hubble diagram to currently unexplored
  redshifts above 2 and up to about 4, from JWST and ELTs
\end{itemize}

In summary the prospects are very exciting. However coordination of
survey execution between facilities is required to maximise the gains;
for example survey fields must be chosen so that they are observable
by both space and ground-based facilities, and scheduling must permit
such coordination. In particular exploitation of the synergies between
JWST and the ELTs and between Euclid/WFIRST and LSST would be
particularly valulable.

\ack{
IH is supported by grants from STFC, OPTICON (EC FP7 grant number
226604) and the UK Space Agency for work on the European ELT and
Euclid projects. The E-ELT/HARMONI simulations in
Section~\ref{sec:eelt} were carried out by Tim Goodsall and
IH. Calculations for JWST observations of $z>2$ SNe
(Section~\ref{sec:jwst}) were carried out by W. Taylor and IH.
Section~\ref{sec:euclid} is based on work undertaken within the Euclid
SN \& transients SWG (part of the Euclid Consortium), and makes use of
calculations performed by P. Astier, S. Spiro, K. Maguire and
J. Guy. Helpful input on the LSST SN search (section~\ref{sec:lsst})
was provided by M. Wood-Vasey.}

\end{document}